\documentclass[12pt]{article}
\pdfoutput=1
\usepackage[numbers,sort&compress]{natbib}
\usepackage{graphicx}
\usepackage{amsmath,graphicx,amssymb,subfigure}
\usepackage{amssymb}
\usepackage{amsmath}
\usepackage{stmaryrd}
\usepackage{enumerate}
\usepackage{isomath}
\usepackage{amsfonts}
\usepackage[all]{xy}
\usepackage[section]{placeins}
\usepackage[colorlinks=true,linktocpage=true,linkcolor=blue,citecolor=blue]{hyperref}
\newcommand{\mt}[1]{\textrm{\tiny #1}}
\newcommand{\be}{\begin{equation}}
\newcommand{\ee}{\end{equation}}
\newcommand{\bea}{\begin{eqnarray}}
\newcommand{\eea}{\end{eqnarray}}
\newcommand{\n}{\nonumber}
\newcommand{\rh}{r_\mt{H}}

\begin{document}

\bibliographystyle{hieeetr}

\pagestyle{plain}
\setcounter{page}{1}

\begin{titlepage}

\begin{center}

\vskip 30mm

{\Large {\bf Linear and quadratic in temperature resistivity from holography  }}

\vskip 0.8 cm

{\bf{Xian-Hui Ge}}$^{1,4,5}$,  {\bf{Yu Tian}}$^{2,4}$, {\bf{Shang-Yu Wu}}$^{3}$, {\bf Shao-Feng Wu}$^{1,4,5}$

$^1${\it{Department of Physics, Shanghai University},}
{\it{Shanghai 200444,  China}} \\
$^2$ {\it School of Physics, University of Chinese Academy of Sciences, Beijing, 100049,  China } \\
$^3${\it  Department of Electrophysics, National Chiao Tung University, Hsinchu 300,  R. of China.}\\
$^4${\it  Shanghai Key Laboratory of High Temperature Superconductors, Shanghai 200444,  China}\\
$^5${\it Shanghai Key Lab for Astrophysics, 100 Guilin Road, 200234 Shanghai,  China}\\
{\sf{gexh@shu.edu.cn}}, {\sf{ytian@ucas.ac.cn}}, {\sf{loganwu@gmail.com}},{\sf{sfwu@shu.edu.cn}}
\medskip

\vspace{5mm}
\vspace{5mm}

\begin{abstract}
We present a new black hole solution in the asymptotic Lifshitz spacetime with a hyperscaling violating factor.
A novel computational method is introduced to compute  the DC thermoelectric conductivities analytically.
We find that both the linear-T and quadratic-T contributions to the resistivity can be realized, indicating that  a more detailed comparison with experimental phenomenology can
be performed in this scenario. 
\end{abstract}
\end{center}
 \noindent
\end{titlepage}
\section{Introduction}
The normal states of high temperature superconductors and heavy fermion compounds  have become one of the most challenging topics
in condensed matter physics. A clear understanding of
the normal-state transport properties of cuprates is considered as a key step towards understanding the pairing
mechanism for high-temperature superconductivity.  There is still a lack of a satisfying explanation of the linear temperature
dependence of resistivity at sufficiently high temperatures in materials such as organic conductors, heavy fermions, Fullerenes, Vanadium Dioxide, and Pnictides.
 In addition, the quadratic temperature dependence of the Hall angle, the violation of Kohler's rule and the divergence of the resistivity anisotropy are those puzzled the theorists for more than two decades \cite{brooks}.

 The transport properties of the normal states of high temperature superconductors are highly anisotropic with a much higher conductivity parallel to $\rm CuO_2$ plane than the perpendicular direction. The in-plane resistivity of hole-doped cuprates shows a systematic evolution with doping. In the underdoped cuprates, the  in-plane resistivity  varies approximately linearly with temperature at high temperature. But as the temperature cools down,
 the  in-plane resistivity deviates downward from linearity, suggestive of a higher power T-dependence. The optimally doped  cuprates are characterized by a linear-T resistivity for the range above the critical temperature $T>T_c$, whilst on the overdoped side, the linear-T relation is replaced by $T^2$-dependence.  On the other hand, the $T^2$-dependence of the Hall angle can be observed in a wide range of doping from underdoped region to overdoped region.

 The AdS/CFT correspondence provides a powerful prescription for calculating transport coefficients of strongly coupled systems by analyzing small perturbations about the black holes that describe the equilibrium state \cite{Hartnoll,McGreevy,Herzog:2009}. Recently, some of us studied conductivity anisotropy holographically in \cite{gws}.
 In \cite{Blake:2014yla}, Blake and Donos attempted to attack the mystery of the linear temperature resistivity and the quadratic temperature Hall angle phenomena by proposing two different relaxation time scales. One central point of their observations is that the Hall angle is only proportional to the momentum dissipation-dominated conductivity i.e. $\theta_{H}\sim B \sigma_{diss}/q $, where $\sigma_{diss}$ is the momentum dissipation conductivity, $B$ is the magnetic field strength and $q$ is related to the charge density. Hence, the temperature dependence of the Hall angle is different from the DC conductivity  because the DC conductivity is decomposed into the sum of a coherent contribution due to momentum relaxation and an incoherent contribution due to intrinsic current relaxation\footnote{As it was clarified in \cite{blaise2,blake15}, it is not proper to say that DC conductivity has one term stemmed from momentum relaxation and the other term from incoherent contribution since it is inconsistent with the known  behavior of the incoherent hydrodynamic DC conductivities. } \cite{blaise2}.
 They further predicted that the resistivity would take the general form $\rho \sim T^2/(\Delta+T)$, where $\Delta$ is a model dependent energy scale. In the low temperature limit $T\ll \Delta$, the resistivity is governed by the Fermi-liquid
 $T^2$ behavior. The $T^2$-dependence of the Hall angle also signifies the Fermi-liquid phenomena. Conversely, in the high temperature limit $T\gg \Delta$, it shows linear resistivity of strange metals.
 In \cite{jianpin}, the authors studied DC electrical and Hall conductivity in the massive Einstein-Maxwell-Dilaton gravity. They found that the linear-T and quadratic-T resistivity can be simultaneously achieved in Lifshitz spacetimes at a dynamical exponent $z=6/5$ and a hyperscaling violating exponent $\theta=8/5$. Other works addressing on the linear-T resistivity and Hall angle can be found in \cite{pal11,pal12,gout11,gout10,oz13,gout14,gout1401,pang,pang2,lucas,karch,hartnoll,Amoretti16} for an incomplete list.

 In this paper, we report our construction of a new asymptotic Lifshitz black hole solution in the Einstein-Maxwell-dilaton-axion model with a hyperscaling violating exponent.
 The solution is supported by two gauge fields and a dilationic scalar, the former playing very different roles. One gauge field is responsible for generating the Lifshitz-like vacuum of
 the background. The other plays a role analogous to that of a standard Maxwell field in asymptotically AdS space.
  The general expressions of transport coefficients are then calculated. When focusing on special cases with $z=1$ in which the metric corresponds to asymptotically AdS space, one can easily achieve a resistivity with two time scales in the asymptotic AdS spacetime. It is well known that in  real materials, the spatial translation invariance is broken and the momentum of charge carriers is not conserved because of the presence of impurities and lattices \cite{donos,Horowitz:201204,Horowitz:2013,Hartnoll:2012,johanna,YiLing:2013,lingprl,Donos:2013eha,YiLing:2014,zhou,Vegh,Davison1,Blake:2013,Blake:20132,WJP:2014,Davison2,
 withers57,kim,cheng,Davison:2014,Gout¨¦raux:2014,gls,gautlett1,gautlett2,gautlett3,kim2,tian, andrade,blaise3,dibakar,Amoretti14,Amoretti15,sun}. In this paper, the translational symmetry breaking is realized through introducing linear-spatial coordinates dependent axions.
  An established means to test whether quasiparticles and thus Landau's Fermi-liquid theory valid, is to compare the thermal conductivity and the electrical conductivity \cite{brooks}. If quasiparticles can be well defined, the Wiedemann-Franz law characterizes the zero temperature value of the Lorenz number $L_0=\pi^2/3\times k^2_B/e^2$, where $k_B$ is Boltzmann's constant and $e$ is the charge of an electron. If in a system $L/L_0$ equals one, we say that Fermi liquid description is exactly satisfied.  On the other hand, $L/L_0>1$ means that there are additional carriers which contribute to the heat current but not to the charge current.
   By contrast,  $L/L_0<1$  at zero temperature implies the breakdown of Landau's Fermi-liquid picture \cite{donos,gls}.   In this paper, all the thermoelectric conductivities  and the Lorenz ratio will be computed in this model. We also would like to check the Wiedemann-Franz law at zero temperature. Although in the holographic setup, the metal has no relationship whatever with real Fermi liquids, the strange metal scaling geometries presented here maybe able to mimic Fermi liquid behavior in transport \cite{boer,tavanfar,allais}.

The structure of this paper is organized as follows. In section 2, we present a new black hole solution in general $(d+2)$-dimensional Lifshitz spacetime. We then calculate the DC electrical conductivity,
thermal conductivity and thermoelectric conductivity in terms of the horizon data in section 3. We develop a new method in calculating the DC transport coefficients. Discussions and conclusions are presented in section 4.

\section{A new black brane solution in Lifsthitz spacetime with linear axion fields and hyperscaling violating factor }
Let us begin with a general action
\begin{equation}\label{action1}
S=\frac{1}{16\pi G_{d+2}}\int d^{d+2} x \sqrt{-g}[R+V(\phi)-\frac{1}{2}(\partial\phi)^2-\frac{1}{4}\sum_{i=1}^n Z_i(\phi) F_{(i)}^2-\frac{1}{2}Y(\phi)\sum\limits_{i}^{d}(\partial \chi_i)^2],
\end{equation}
where we have used the notation $Z_i= e^{\lambda_i \phi}$ and $Y(\phi)= e^{-\lambda_2\phi}$. Note that $R$ is the Ricci scalar and $\chi_i$ is a collection of $d-$massless linear axions.
 The action consists of Einstein gravity, axion fields, and $U(1)$ gauge fields and  a dilaton field. For simplicity, we only consider two $U(1)$ gauge $F^{(1)}_{rt}$ and $F^{(2)}_{rt}$ in which the first gauge field plays the role of an auxiliary field, making the geometry asymptotic Lifshitz, and the second gauge field makes the black hole charged,  playing a role analogous to that of a standard Maxwell field in asymptotically AdS space.

Solving the equations of motion, we are able to obtain a spacetime which is asymptotically Lifshitz and hyperscaling violated.
 The action yields a Lifshitz black brane solution with a hyperscaling violating factor
\bea\label{metric}
&& ds^2=r^{-\frac{2\theta}{d}}\bigg(-{{r^{2z}}}f(r)dt^2+\frac{dr^2}{r^2 f(r)}+{ {r^2}}d\vec{x}^2_{d}\bigg),\\
 &&f(r)=1-{\frac{m}{r^{d+z-\theta}}}+{\frac{Q^2}{r^{2(d+z-\theta-1)}}}-{\frac{ \beta^2}{r^{2z-2\theta/d}}},\label{fr}\\
 &&F_{(1)rt}=Q_1\sqrt{2(z-1)(z+d-\theta)}r^{d+z-\theta-1},\\
 &&F_{(2)rt}=Q_2 \sqrt{2(d-\theta)(z-\theta+d-2)}r^{-(d+z-\theta-1)},\\
 &&\lambda_1=-\frac{2d-2\theta+\frac{2\theta}{d}}{\sqrt{2(d-\theta)(z-1-\theta/d)}},\\
 &&\lambda_2=\sqrt{{2}{\frac{z-1-\theta/d}{d-\theta}}},\\
  &&e^{\phi}=r^{\sqrt{2(d-\theta)(z-1-\theta/d)}},~~~V(\phi)=(z+d-\theta-1)(z+d-\theta)r^{2\theta/d},\\
  && \chi_i= \beta_{ia} x^a, ~~~\beta^2_0=\frac{1}{d}\sum^{d}_{i=1}\overrightarrow{\beta}_a \cdot \overrightarrow{\beta}_a,~~~\overrightarrow{\beta}_a \cdot \overrightarrow{\beta}_b= \beta^2_0 \delta_{ab} ~~~{\rm for}~~~ i\in \{1,d\}.
 \eea
 where $\beta^2={\frac{d^2 \beta^2_0}{2(d-\theta)(d^2+2\theta-(z+\theta)d)}}$.
 This solution is Lifshitz-like even in the UV.
  When the dynamical exponent $z=1$, we recover the normal AdS black hole geometry because $F_{(1)rt}=0$. The black hole solution can return to the result given in \cite{alisha} and \cite{Tarrio} under  the condition of $\beta=0$ and $\theta=0$, respectively.
  The transport coefficients have been studied in \cite{kuang}. We emphasize that the choice of couplings $Y(\phi)$ and $Z_i(\phi)$ is our choice here and we believe that different choices
  of coupling would leads to different power scalings of the transport. Intriguingly, in a later paper, the exact solution presented here was found again by the authors  of \cite{Hlu}.
The event horizon locates at $r=\rh$ satisfying the relation $f(\rh)=0$. We can express the mass $m$ in terms of $\rh$
\bea
m=\rh^{d+z-\theta}+Q^2_2 \rh^{2-d-z+\theta}-\beta^2 \rh^{d-z-\theta+2\theta/d}.
\eea
By further introducing a coordinate $z=\rh/r$, we can recast $f(r)$ as
\be
f(z)=1-z^{d+z-\theta}+\frac{Q^2_2}{\rh^{2(d+z-\theta-1)}}\bigg[z^{2(d+z-\theta-1)}-z^{d+z-\theta}\bigg]
+\frac{\beta^2}{\rh^{2z-2\theta/d}}\bigg[z^{d+z-\theta}-z^{2z-2\theta/d}\bigg].
\ee
The corresponding Hawking temperature is given by
\be
T=\frac{(d+z-\theta)\rh^z}{4\pi}\bigg[1-\frac{d+z-\theta-2}{d+z-\theta}Q^2_2\rh^{-2(d+z-\theta-1)}-\frac{d^2+2\theta-(z+\theta)d}{d(d+z-\theta)}\rh^{2\theta/d-2z}\beta^2\bigg].
\ee
The entropy density is given by $s=\rh^{d-\theta}/4G$. The specific heat of this black hole can be evaluated via $c=T(\partial s/\partial T)_{Q,\beta}$. We find that the thermodynamical stability  and the positiveness of the specific heat require $\theta<d$.
The near horizon geometry can be evaluated by introducing two new coordinates $u$ and $\tau$:
\be
r-\rh=\frac{\epsilon\rh^2}{l^2 u},~~~t=\frac{\tau}{\epsilon \rh^{z-1}}. \nonumber\\
\ee
We can see that at zero temperature $T=0$, the solution near the horizon develops an $AdS_2\times R^{d-1}$ geometry. The near horizon geometry is defined by the limit $\epsilon\rightarrow 0$:
\be
ds^2=\rh^{-2-\frac{2\theta}{d}} \bigg(\frac{-d\tau^2+du^2}{l^2 u^2}\bigg)+\rh^{2-\frac{2\theta}{d}}d\vec{x}_d.
\ee
The effective $AdS_2$ radius is given by:
\bea
l^2_{ads_2}&=&\frac{\rh^{-2-\frac{2\theta}{d}}}{l^2} ,\\
l^2&=&(d-1)(d-\theta)(d+z-\theta-2)Q^2_2 \rh^{2(\theta-d-z)}/d+(d+z-\theta)(dz-\theta)\rh^{-2}/d .
\eea
We observe that even in the absence of the $U(1)$ gauge field, the black brane could still be extremal with near horizon of $AdS_2$ as we just demonstrated.
It means that at low temperature the theory flows to an IR fixed point in the presence of the linear axion fields.\\

$\bullet~~~$ Black hole solution at $(d+z-\theta-2)=0$ \\

One may notice that as  $(d+z-\theta-2)\rightarrow 0$, $Q_2$ and $f(r)$ appear to diverge.   At well-defined solution can be achieved in an alternative form:
\bea
f(r)&=&1-{\frac{m}{r^{d+z-\theta}}}-\frac{q^2_2\ln r}{2(d-\theta)r^{d+z-\theta}}-{\frac{ \beta^2}{r^{2z-2\theta/d}}},\label{flog}\\
&=&1-{\frac{m}{r^{2}}}-\frac{q^2_2\ln r}{2(2-z)r^{2}}-{\frac{ \beta^2}{r^{2z-2\theta/d}}},\nonumber\\
F_{(2)rt}&=&q_2 r^{-1}, \label{frt}
\eea
where $M$ and $q_2=Q_2 \sqrt{2(d-\theta)(z-\theta+d-2)}$ are finite physical parameters without divergence as $(d+z-\theta-2)\rightarrow 0$. A careful examination of (\ref{flog}) and (\ref{frt}) reveals that they satisfy the corresponding Einstein equation and Maxwell equation. We can express $f(r)$ in terms of the event horizon radius
\be
f(r)=1-\frac{\rh^2}{r^2}+\frac{q^2_2}{2r^2(2-z)}\ln \frac{\rh}{r}-\frac{\beta^2}{\rh^{2 z-{2 \theta }/{d}}}\bigg(\frac{\rh^2}{r^2}-\frac{\rh^{2 z-{2 \theta }/{d}}}{r^{2 z-{2 \theta }/{d}}}\bigg).
\ee
The Hawking temperature is given by
\be
T=\frac{\rh^z}{2\pi}\bigg(1-\frac{q^2_2}{4(2-z)\rh^2}-\frac{\beta^2 (d+\theta-dz)}{d \rh^{2 z-{2 \theta }/{d}}}\bigg).
\ee
\section{DC transport coefficients}
Firstly, we would like to introduce a new method by taking advantage of the matrix theory and the equations of motion, which maybe called the matrix method, to calculate the DC electrical and thermoelectric conductivities. The standard calculational method will be presented in section 3.2 as a consistent check and the thermal conductivity will be computed.   In what follows, we work in the special case with $d=2$. Later, we will extend our discussions to more general conditions.

\subsection{DC electrical and thermoelectric conductivities}

For simplicity, we rewrite  the metric  in $d=2$ dimensional spacetime as
\be
ds^2=-g_{tt}dt^2+g_{rr}dr^2+g_{xx}dx^2+g_{xx}dy^2.
\ee
For the purpose of computing the electrical conductivity, we consider the linear perturbations of the form
\bea
&&A_{(1)x}=a_1(r)e^{-i\omega t},\\
&&A_{(2)x}=a_2(r)e^{-i\omega t},\\
&&h_{tx}=h_{tx}(r) e^{-i\omega t},\\
&&\chi_1=\beta x+ \bar{\chi}_1 (r) e^{-i\omega t},
\eea
and let the other metric and gauge perturbations vanishing. Since we choose the conductivity along the $x-$ direction, it is consistent
to set all scalar fluctuations to be vanished except for the one with the linear piece along
the direction $x$.  We can arbitrarily denote this scalar by $\chi$ and write $\chi=\beta x+ \bar{\chi} (r) e^{-i\omega t}$.
The equation of motion for the linear perturbation can be obtained as
\bea
&&\bigg(\sqrt{\frac{g_{tt}}{g_{rr}}}Z_2 a'_1\bigg)'+\frac{A'_{(1)t} Z_1g_{xx}}{\sqrt{g_{tt}g_{rr}}}\bigg(g^{xx}h_{tx}\bigg)'+\omega^2 \sqrt{\frac{g_{rr}}{g_{tt}}} Z_2 a_1=0,\label{a1}\\
&&\bigg(\sqrt{\frac{g_{tt}}{g_{rr}}}Z_2 a'_2\bigg)'+\frac{A'_{(2)t} Z_2g_{xx}}{\sqrt{g_{tt}g_{rr}}}\bigg(g^{xx}h_{tx}\bigg)'+\omega^2 \sqrt{\frac{g_{rr}}{g_{tt}}} Z_2 a_2=0,\label{jx}\\
&&\bigg(\sqrt{\frac{g_{tt}}{g_{rr}}}g_{xx}Z_2  \bar{\chi}'\bigg)'+\omega^2\sqrt{\frac{g_{rr}}{g_{tt}}}g_{xx}Z_2\bar{\chi}-i\omega \beta^2 Z_2\sqrt{\frac{g_{rr}}{g_{tt}}}h_{tx}=0,~~~\label{htx}\\
&&\bigg(g^{xx}h_{tx}\bigg)'+\frac{i \bar{\chi}' g_{tt}}{\omega Z_2}+Z_1 A'_{(1)t} a_1+Z_2 A'_{(2)t} a_2=0,\label{con}\\
&&\bigg(\frac{g^2_{xx}}{\sqrt{g_{rr}g_{tt}}}h'_{tx}\bigg)'-q_1 a'_1-q_2 a'_2-\beta^2 g_{xx} Y\sqrt{\frac{g_{rr}}{g_{tt}}}h_{tx}
-i\omega g_{xx}Y \sqrt{\frac{g_{rr}}{g_{tt}}} \bar{\chi}=0,\label{eqhtxp}
\eea
where the prime denotes a derivative with respect to $r$.  Note that the derivative of the scalar potential is given by $A'_{(1)t}=-\frac{q_1}{Z_1(\phi)}\frac{\sqrt{g_{tt}g_{rr}}}{g_{xx}}$ and
$A'_{(2)t}=-\frac{q_2}{Z_2(\phi)}\frac{\sqrt{g_{tt}g_{rr}}}{g_{xx}}$, where $q_1=Q_1\sqrt{2(z-1)(z+d-\theta)}$ and $q_2=Q_2 \sqrt{2(d-\theta)(z-\theta+d-2)}$. Equation (\ref{con}) is a constrained equation, which implies that the linear perturbations $a_1$, $a_2$, $h_{tx}$ and $\bar{\chi}$ are not all linearly independent.

After introducing $\tilde{\chi}=f r^{z-5}\bar{\chi}'/(i\omega)$ and eliminate $h_{tx}$, we are able to  rewrite the equations (\ref{a1}-\ref{con}) in a more explicit form
\begin{eqnarray}
(r^{z-3+\theta}fa_{1}^{\prime})^{\prime} & = & A_1 a_{1}+B_1 a_2+C_1 \tilde{\chi},\label{ma1}\\
(r^{3z-1-\theta}fa_{2}^{\prime})^{\prime} & = & A_2 a_{1}+B_2 a_2+C_2 \tilde{\chi},\label{ma2}\\
(r^{3(z-1)}f\tilde{\chi}^{\prime})^{\prime} & = & A_3 a_{1}+B_3 a_2+C_3 \tilde{\chi}, \label{ma3}
\end{eqnarray}
where
\[
A_1=\bigg(\frac{q^2_1}{r^{5-z-\theta}}-\frac{\omega^2}{r^{5+z-\theta}f}\bigg),\qquad A_2=B_1=\frac{q_1 q_2}{r^{5-z-\theta}},
\]
\[
B_2=\bigg(\frac{q^2_2}{r^{5-z-\theta}}-\frac{\omega^2}{r^{3-z+\theta}f}\bigg),\qquad B_3=C_2=-\frac{\beta q_2}{r^{5-z-\theta}},
\]
\[C_3=\bigg(\frac{\beta^2}{r^{5-z-\theta}}-\frac{\omega^2}{r^{5-z}f}\bigg), \qquad A_3=C_1=-\frac{\beta q_1}{r^{5-z-\theta}}.
\]
We notice that the combination $(\ref{ma1})+(\ref{ma3})\times q_1/\beta$ and $(\ref{ma2})+(\ref{ma3})\times q_2/\beta$ leads to
\begin{eqnarray}
\bigg(r^{z-3+\theta}fa_{1}^{\prime}+\frac{q_1}{\beta}r^{3(z-1)}f\tilde{\chi}^{\prime}\bigg)^{\prime} & = & 0,\label{massless1}\\
\bigg(r^{3z-1-\theta}fa_{2}^{\prime}+\frac{q_2}{\beta}r^{3(z-1)}f\tilde{\chi}^{\prime}\bigg)^{\prime} & = &0. \label{massless2}
\end{eqnarray}
A massless mode can be extracted from (\ref{massless1}) and (\ref{massless2}). From the membrane paradigm approach \cite{liu} we know that the realization of the currents in the boundary theory can be identified with radially independent quantities in the bulk.
From (\ref{a1}) to (\ref{con}), one can easily find that the equivalent expressions of the conserved electric currents in the zero frequency limit
read
\bea
J_1=-r^{z-3+\theta}fa_{1}^{\prime}+q_1 r^{\theta-2} h_{tx},\label{jx1}\\
J_2=-r^{3z-1-\theta}fa_{2}^{\prime}+q_2 r^{\theta-2} h_{tx}.\label{jx2}
\eea
The DC conductivity is the zero frequency limit of the optical conductivity
\be
\sigma_{ij}^{DC}=\lim_{\omega\rightarrow 0}\sigma_{ij}^{DC}(\omega)=\lim_{\omega\rightarrow 0} \frac{\partial J_i(\omega)}{\partial E_j (\omega)}
\ee
The DC conductivity can be evaluated at the horizon whenever we have massless mode since it does not evolve between the horizon and the boundary \cite{withers57}.
Then let us define a matrix $\tilde{\sigma}$ from
\[
\left\llbracket\begin{array}{c}
r^{z-3+\theta}fa_{1}^{\prime}\\
r^{3z-1-\theta}fa_{2}^{\prime}\\
r^{3(z-1)}f\tilde{\chi}^{\prime}
\end{array}\right\rrbracket=\tilde{\sigma}\left\llbracket\begin{array}{c}
i\omega a_{1}\\
i\omega a_{2}\\
i\omega \tilde{\chi}
\end{array}\right\rrbracket,
\]
where the special notation $\llbracket ...\rrbracket$ should be considered as a square matrix which is introduced  for convenience, for example
\bea\left \llbracket \begin{array}{c}
 a_{1}\\
 a_{2}\\
 \tilde{\chi}
\end{array}\right \rrbracket\equiv
 \left(\begin{array}{ccc}
 a_{1} & a^{(2)}_{1} & a^{(3)}_{1}\\
a_{1} & a^{(2)}_{2} & a^{(3)}_{2}\\
\tilde{\chi} &\tilde{\chi}^{(2)} & \tilde{\chi}^{(3)}
\end{array}\right),
\eea
in which $a^{(i)}_{1}$, $a^{(i)}_{2}$ and $\tilde{\chi}^{(i)}$  are linearly independent sources, introduced to guarantee the source term invertible. After inverting the components in $\llbracket ...\rrbracket$, $a^{(i)}_{1}$, $a^{(i)}_{2}$ and $\tilde{\chi}^{(i)}$ will be not important in further calculations and it is better for us to hide them in $\llbracket ...\rrbracket$.
We emphasize that the matrix $\tilde{\sigma}$ is not the exact conductivity tensor of the system as we can see below.
We take the derivative of  $\tilde{\sigma}$ and obtain
\begin{eqnarray}
\tilde{\sigma}^{\prime} & = & \left\llbracket\begin{array}{c}
r^{z-3+\theta}fa_{1}^{\prime}\\
r^{3z-1-\theta}fa_{2}^{\prime}\\
r^{3(z-1)}f\tilde{\chi}^{\prime}
\end{array}\right\rrbracket^{\prime}\left\llbracket\begin{array}{c}
i\omega a_{1}\\
i\omega a_{2}\\
i\omega \tilde{\chi}
\end{array}\right\rrbracket^{-1}-i\omega\tilde{\sigma}\left\llbracket\begin{array}{c}
 a_{1}\\
 a_{2}\\
 \tilde{\chi}
\end{array}\right\rrbracket^{\prime}\left\llbracket\begin{array}{c}
i\omega a_{1}\\
i\omega a_{2}\\
i\omega \tilde{\chi}
\end{array}\right\rrbracket^{-1}\n\\
 & = & \left\llbracket\begin{array}{c}
 A_1 a_{1}+B_1 a_2+C_1 \tilde{\chi}\\
 A_2 a_{1}+B_2 a_2+C_2 \tilde{\chi}\\
  A_3 a_{1}+B_3 a_2+C_3 \tilde{\chi}
 \end{array}\right\rrbracket\left\llbracket\begin{array}{c}
 i\omega a_{1}\\
 i\omega a_{2}\\
 i\omega \tilde{\chi}
\end{array}\right\rrbracket^{-1}-i\omega\tilde{\sigma}\left\llbracket\begin{array}{c}
 a'_{1}\\
 a'_{2}\\
 \tilde{\chi}'
\end{array}\right\rrbracket\left\llbracket\begin{array}{c}
i\omega a_{1}\\
i\omega a_{2}\\
i\omega \tilde{\chi}
\end{array}\right\rrbracket^{-1}\n\\
 & = & \frac{1}{i\omega}\left(\begin{array}{ccc}
A_1 & B_1& C_1\\
A_2 & B_2& C_2\\
A_3 & B_3& C_3
\end{array}\right)-i\omega\tilde{\sigma}\left(\begin{array}{ccc}
(r^{z-3+\theta}f)^{-1}& 0 & 0\\
0 & (r^{3z-1-\theta}f)^{-1}&0\\
0 & 0 & (r^{3(z-1)}f)^{-1}
\end{array}\right)\tilde{\sigma}.\label{eqflow}
\end{eqnarray}
The prime denotes the derivative with respect of $r$.
The advantage of this method is that it reduce second order ordinary differential equations to non-linear first order ordinary differential equations.
Multiplying both sides of equation (\ref{eqflow}) with $f$, we obtain  
\[
f\tilde{\sigma}^{\prime}=\frac{f}{i\omega}\left(\begin{array}{ccc}
A_1 & B_1& C_1\\
A_2 & B_2& C_2\\
A_3 & B_3& C_3
\end{array}\right)-i\omega\tilde{\sigma}\left(\begin{array}{ccc}
r^{3-\theta-z} & 0 &0\\
0 & r^{1+\theta-3z}&0\\
0 & 0& r^{3-3z}
\end{array}\right)\tilde{\sigma}.
\]
At the event horizon $f(\rh)=0$ and $\tilde{\sigma}^{\prime}$ is finite.  So the above equation reduces to
\[
0=\left(\begin{array}{ccc}
\rh^{\theta-z-5} & 0 &0\\
0 & \rh^{z-\theta-3}&0\\
0 & 0& \rh^{z-5}
\end{array}\right)-\tilde{\sigma}_0\left(\begin{array}{ccc}
\rh^{3-\theta-z} & 0 &0\\
0 & \rh^{1+\theta-3z}&0\\
0 & 0& \rh^{3-3z}
\end{array}\right)\tilde{\sigma}_0.
\]
The regularity condition at the event horizon yields
\[
 \tilde{\sigma}_0=\left(\begin{array}{ccc}
\rh^{-4-\theta} & 0&0\\
0 & \rh^{2z-2+\theta}&0\\
0 & 0& \rh^{2z-4}
\end{array}\right).
\]
From the definition of the matrix $\tilde{\sigma}$, we obtain the boundary condition at the event horizon
\begin{eqnarray}
fa_{1}^{\prime} & \to & i\omega \rh^{-z-1}a_{1}\Big|_{\rh},\label{eq:a_x1}\\
fa_{2}^{\prime} & \to & i\omega \rh^{-z-1}a_{2}\Big|_{\rh},\label{eq:a_x}\\
f\tilde{\chi}^{\prime} & \to & i\omega \rh^{-z-1}\tilde{\chi}\Big|_{\rh}\label{eq:phi}.
\end{eqnarray}
Considering the above relation (\ref{eq:a_x1})-(\ref{eq:phi}),  we then can impose  the regularity condition at the horizon from equation (\ref{eqhtxp}) and obtain
\begin{equation}\label{chicon}
h_{tx}\big|_{r=\rh}=\bigg(-i \omega \frac{q_1}{\beta^2 Y}a_1-i \omega \frac{q_2}{\beta^2 Y}a_2-i\omega \frac{\bar{\chi}}{\beta^2 Y}\bigg)\bigg|_{r=\rh}.
\end{equation}
The last term in the right hand of (\ref{chicon}) will be dropped out in the following calculation since it does not contribute to the transport.
Further utilizing (\ref{eq:a_x1}), (\ref{eq:a_x}), (\ref{jx1}) and (\ref{jx2}), we can determine the value of currents
\begin{eqnarray}
J_1&=&-\bigg(\rh^{\theta-4}+\frac{q^2_1}{\beta^2}\rh^{2z-4}\bigg)i\omega a_1-\frac{q_1 q_2}{\beta^2}\rh^{2z-4}i\omega a_2,\label{jcurrent1}\\
J_2&=&-\bigg(\rh^{2z-2-\theta}+\frac{q^2_2}{\beta^2}\rh^{2z-4}\bigg)i\omega a_2-\frac{q_1 q_2}{\beta^2}\rh^{2z-4}i\omega a_1.\label{jcurrent2}
\end{eqnarray}
The DC electric conductivity can be computed via $\sigma_{ij}=\frac{\partial J_i}{\partial E_j}$, where $E_j=-i\omega a_j$. Finally we obtain
\begin{eqnarray}
\sigma_{11}&=&\rh^{\theta-4}+\frac{q^2_1}{\beta^2}\rh^{2z-4},~~~\sigma_{12}=\frac{q_1 q_2}{\beta^2}\rh^{2z-4},\nonumber\\
\sigma_{21}&=&\sigma_{12},~~~~~~\sigma_{22}=\rh^{2z-2-\theta}+\frac{q^2_2}{\beta^2}\rh^{2z-4}.\label{contensor1}
\end{eqnarray}
This result is consistent with \cite{Hlu}. The physical interpretation of the DC conductivity tensor obtained here is somehow subtle: we consider electric perturbations only along the $x-$direction but obtain a  $2\times 2$ conductivity matrix with non-vanishing off-diagonal components. We also observe that taking $z\rightarrow 1$, $\theta\rightarrow 0$ and then $q_1\rightarrow 0$, but the quantity $\sigma_{11}=\rh^{-4}$ is not vanishing. However, if we set $z=1$ and $\theta=0$ from the very beginning in the action (\ref{action1}),
the auxiliary gauge field $F_{(1)rt}$  naturally does not appear and the black hole solution is the Reissner-Nordstr$\rm\ddot{o}$m-AdS metric with vanishing $\sigma_{11}$ and $\sigma_{12}$. So we have
a discontinuity in the $z\rightarrow 1$,  $\theta\rightarrow 0$  and $q_1 \rightarrow 0$ limit. This means that once we change the asymptotic structure from an AdS to a Lifshitz one and turn on the perturbation $\delta A_{(1)x}$, it could not have a continuous limit back to the perturbation considered in the Reissner-Nordstr$\rm\ddot{o}$m-AdS spacetime by simply taking  $z\rightarrow 1$, $\theta \rightarrow 0$ and $q_1 \rightarrow 0$ limit.

The original purpose of introducing the auxiliary $U(1)$ gauge field $F_{(1)rt}$ is to construct the Lifshitz-like nature of the vacuum. One may notice that not only $A_{(1)t}$ , but also $a_1$ diverges in the asymptotic $r\rightarrow \infty$ regime:
\be
a_1=a_{10}+\frac{a_{20}}{r^{z-4+\theta}},
\ee
where the second term diverges when $z-4+\theta< 0$ at the infinite boundary. So that we must impose the regular condition  $a_{20}=0$. That is to say, $a_1$ does not introduce a charge current on the asymptotic boundary. In this sense, we should set the boundary condition $J_1=0$. From (\ref{jcurrent1}),  (\ref{jcurrent2}) and $\sigma_{ij}=\frac{\partial J_i}{\partial E_j}$, we obtain
\be \label{mainresult}
\sigma_{DC}=\rh^{2z-2-\theta}+\frac{q^2_2}{(\beta^2+q^2_1 \rh^{2z-\theta})}\rh^{2z-4}.
\ee
This is a very intriguing result because (\ref{mainresult}) means that even without translational symmetry breaking, finite DC electric conductivity can still be realized because of the presence of  the auxiliary $U(1)$ charge $q_1$ \cite{sonner}. By embedding the Lifshitz solution in AdS, the divergence encountered here is no longer a problem since an AdS embedding modifies the UV properties without affecting the horizon behavior. However, it is not of our purpose to realize such an AdS embedding in this paper.

\begin{figure}
\center{
\includegraphics[scale=0.4]{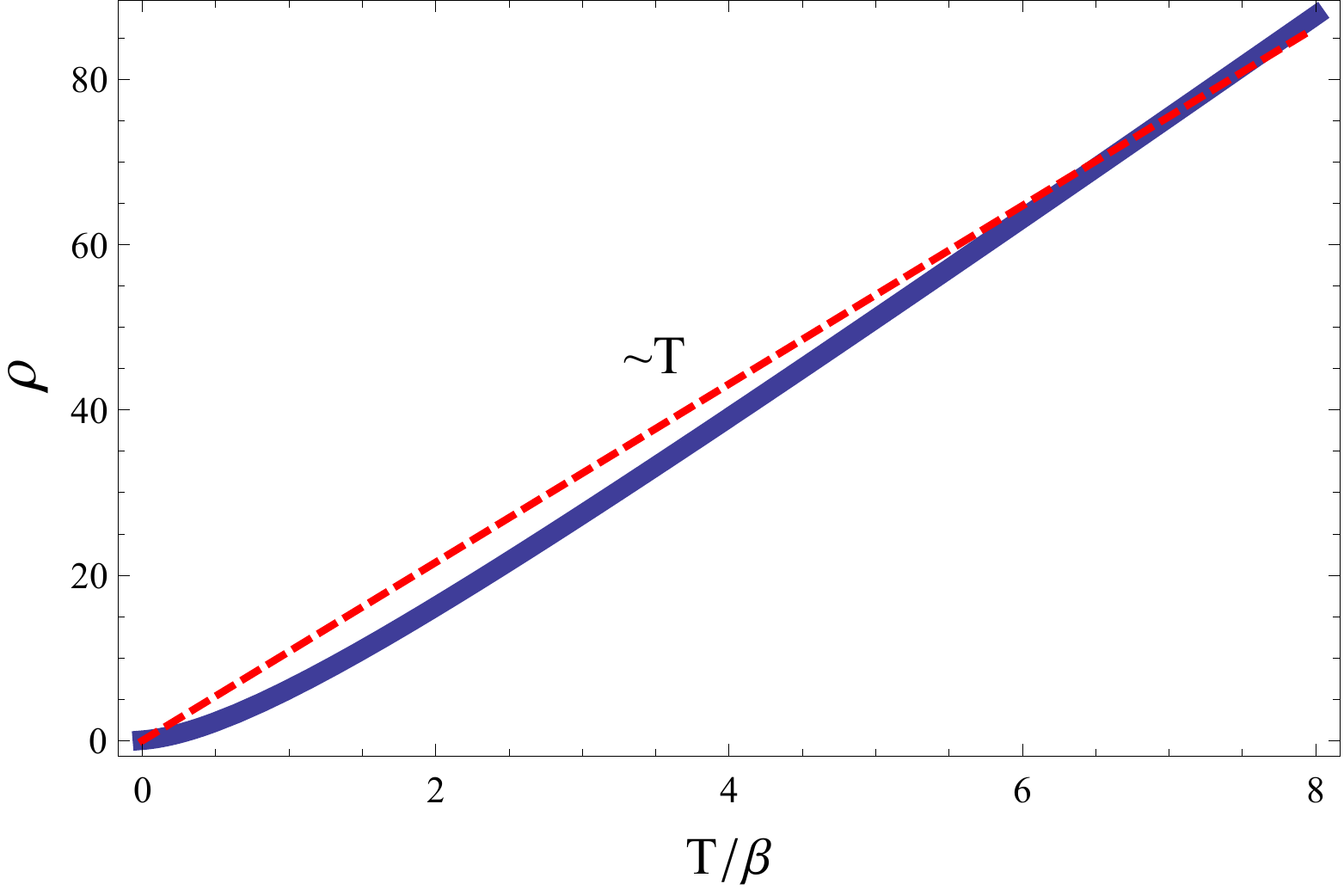}\hspace{0.3cm}
\includegraphics[scale=0.4]{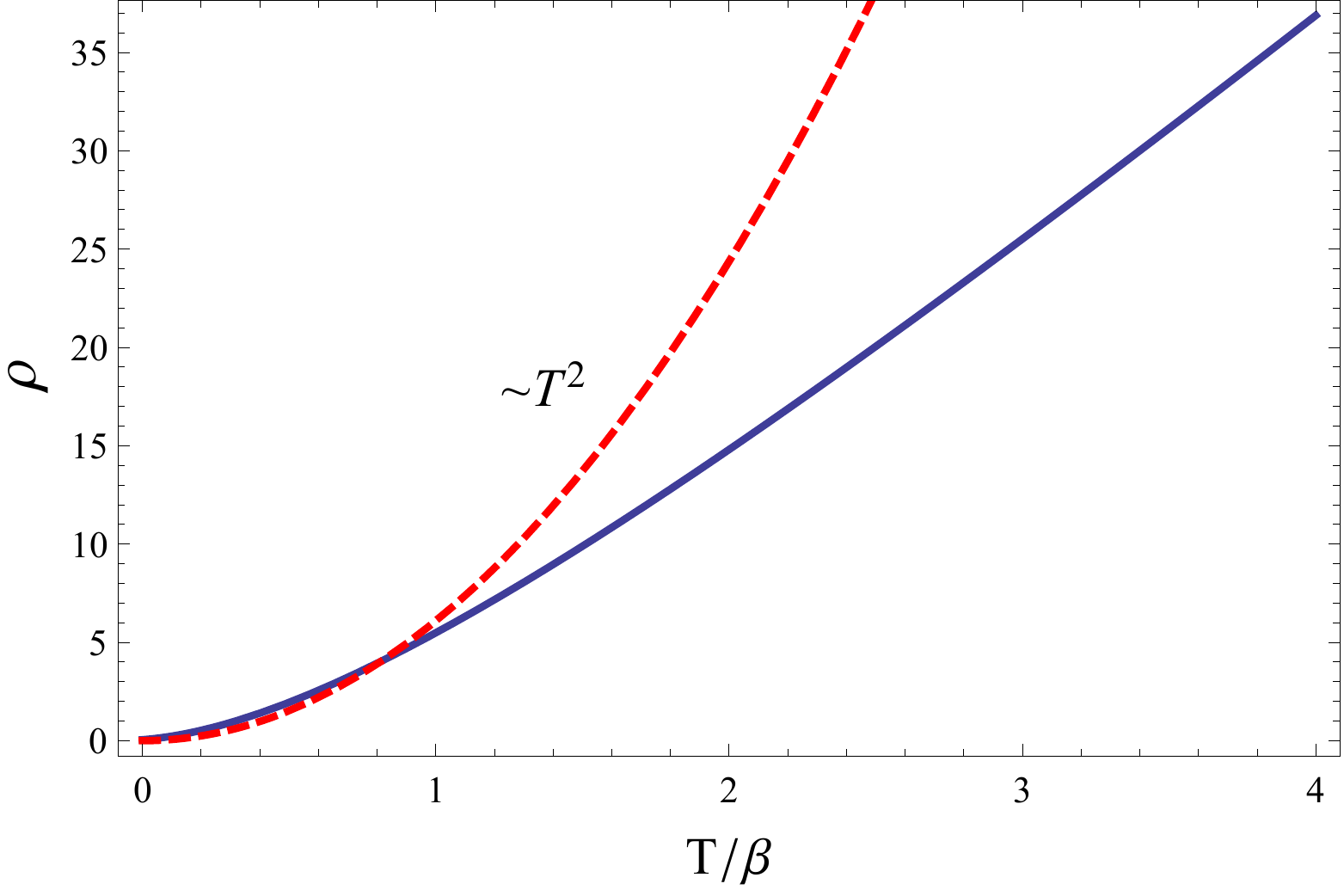}
\caption{\label{solu} The resistivity as a function of temperature.  (Left) The resistivity shows linear-T behavior at higher temperature with $q^2_2/\beta^4=8$.  (Right)
The resistivity shows quadratic-T behavior at lower temperature with $q^2_2/\beta^4=10$. The dashed red lines correspond to fitting functions
$\rho\sim 10.78 T/\beta$ and $\rho\sim 6.09 T^2/\beta^2$, respectively.} }\label{rhoT}
\end{figure}

Another interesting  situation is  the case without
 translational invariance breaking (i.e. $\beta=0$). We also arrive at a finite conductivity
 \be
\sigma_{DC}=\rh^{2z-2-\theta}+\frac{q^2_2}{q^2_1 \rh^{2z-\theta}}\rh^{\theta-4}.
\ee
The linear and quadratic in temperature resistivity can be reached via $z=6/5$ and $\theta=8/5$. Note that these are the exact exponents given in \cite{jianpin}. This feature of the construction of a finite conductivity without the need to break translational invariance has been reported and explained by Sonner in \cite{sonner}. Throughout this paper, we mainly consider the situation
with $J_1=0$, because it is mathematically inconsistent to turn off $a_1$. However, it is also physically unclear of the boundary correspondence of the source $a_1$ because the
auxiliary gauge field is only introduced to realize Lifshitz-like vacuum. Therefore, it is consistent to set $J_1=0$.

Considering two gauge fields resulting a $2\times 2$ electric conductivity matrix, one naturally expects that the thermoelectric conductivity  has more than one component.
One may notice the equation of motion for $h_{tx}$ at zero frequency is given by
\bea
h''_{tx}&-&\frac{1}{2}\bigg(\frac{g'_{rr}}{g_{rr}}+\frac{g'_{tt}}{g_{tt}}\bigg)h'_{tx}+\bigg(\frac{g'_{rr}g'_{tt}}{2 g_{rr}g_{tt}}+\frac{g'^2_{tt}}{2 g^2_{tt}}-\frac{g''_{tt}}{g_{tt}}+\frac{Z_2 A'^2_{(2)t}}{g_{tt}}+\frac{Z_1 A'^2_{(1)t}}{g_{tt}}\bigg)h_{tx}\nonumber\\&+&Z_2 A'_{(2)t}a'_2+Z_1 A'_{(1)t}a'_1=0.
\eea
Clearly, the vector type of perturbations $h_{tx}$ is coupled to $a_1$ and $a_2$.
Together with equations of motion of the Maxwell fields to the linear order, we can write down a radially conserved heat current
\be
\mathcal{Q}=\sqrt{\frac{g_{tt}}{g_{rr}}}\bigg(-g^{tt}h_{tx}\partial_r g_{tt}+h'_{tx}\bigg)-A_{(1)t} J_1-A_{(2)t} J_2.
\ee
After imposing the regularity condition at the  event horizon, that is to say
\be
h_{tx}(r=\rh)=\bigg(-i \omega \frac{q_1}{\beta^2 Y}a_1-i \omega \frac{q_2}{\beta^2 Y}a_2+...)\bigg|_{r=\rh},
\ee
 we  can simply evaluate the conserved heat current at the event horizon
\be
\mathcal{Q}=-\frac{4\pi T i \omega \rh^{2z-2-\theta}}{\beta^2 }(q_1 a_1+q_2 a_2)\bigg|_{r=\rh}.
\ee
We have used the boundary condition $A_{(1)t}(\rh)=A_{(2)t}(\rh)=0$. The thermoelectric conductivity can be obtained at the event
horizon $r=\rh$ by using the expression $\bar{\alpha}_{i}=\frac{\partial \mathcal{Q}}{T \partial E_i}$. We finally obtain
\bea
&&\bar{\alpha}_1=\frac{\partial \mathcal{Q}}{T \partial E_1}=\frac{4\pi q_1}{\beta^2 }\rh^{2z-2-\theta},\\
&&\bar{\alpha}_2=\frac{\partial \mathcal{Q}}{T \partial E_2}=\frac{4\pi q_2}{\beta^2 }\rh^{2z-2-\theta}.
\eea
There are no off-diagonal components of the thermoelectric coefficient as can be seen above. Both components obey the same temperature scaling. If one turns on magnetic field, the off-diagonal components of the thermoelectric conductivity can be observed.

$\bullet~~~$ Special case: $z=1$, $\theta=1$ and $J_1=0$ \\
For the case $\theta=1$, $z=1$ and thus $q_1=0$, the temperature is given by
\be
T=\frac{\rh}{2\pi}\bigg(1-\frac{q^2_2}{4\rh^2}-\frac{\beta^2}{2\rh}\bigg).
\ee
In this case, the entropy density $s=\rh/4G$ is proportional to the temperature in the small $q$ and $\beta$ limit.
 We find that the DC electric conductivity (\ref{mainresult}) behaves as
\be \label{dcsig}
\sigma_{DC}=\frac{1}{\rh}+\frac{q^2_2}{\beta^2 \rh^2}
 \sim \frac{1}{2\pi T}+\frac{q^2_2}{4\pi^2 \beta^2 T^2},
\ee
where we have use the large horizon radius approximation $\rh \sim 2 \pi T$.
The resistivity in the small $\beta$ limit can be expressed as
\be \label{rho1}
\rho \approx \frac{4 \beta^2 \pi^2 T^2}{q^2_2+2 \beta^2 \pi T}=\frac{\tilde{T}^2}{\tilde{T}+\Delta},
\ee
where $\tilde{T}=2\pi T$ and $\Delta=q^2_2/\beta^2$. Equation (\ref{rho1}) shows us that for $\tilde{T}\gg \Delta$, the resistivity is dominated by the linear-T behavior, while $\tilde{T} \ll \Delta$, the system obeys the Fermi-liquid like law.
As a demonstration, we plot the resistivity as a function of temperature in figure \ref{rhoT}. In the higher temperature regime, the resistivity shows linear in temperature dependence, analogous to the experimental behavior of bad metals. In the low temperature regime, the resistivity varies as $T^2$, retaining Landau's Fermi-liquid description, although the quasiparticle picture is not well defined here. One can also understand  equation (\ref{dcsig}) as follows: for small $\beta$ but fixed temperature $T$ and charge density $q_2$, (\ref{rho1}) shows Fermi-liquid-like property, while large $\beta$ results in strange metal behavior.
 At zero temperature, the DC conductivity becomes
\be
\sigma_{DC}=\frac{4}{\beta^2}+\frac{\beta^2}{q^2_2}-\frac{\sqrt{\beta^4+4q^2_2}}{q^2_2}.
\ee
This equation implies that as the disorder goes to zero, the system becomes an ideal metal with infinite DC conductivity, while $\beta\rightarrow\infty$ the ground state is an insulator.

\subsection{DC Thermal and thermoelectric conductivities}
In irreversible thermodynamics, the  dissipative properties of a system are closely related to the entropy production in a unit time
\be
\frac{d s}{dt}=\sum_{i} \mathcal{T}_i \mathcal{X}_i,
\ee
where $\mathcal{X}_i$ is the thermal force which is determined by the gradients of energy, temperature, chemical potential etc.  $\mathcal{T}_i$ denotes the current driven by $\mathcal{X}_i$ which can be written in the linear approximation as
\be
\mathcal{T}_i=\sum_{j} \mathcal{L}_{ij} \mathcal{X}_i,
\ee
where $\mathcal{L}_{ij}$ represent the transport coefficients. We can see that both the thermal force $\mathcal{X}_i$ and the transport coefficients $\mathcal{L}_{ij}$ contribute to the entropy production rate. The thermal force
represents the external factor describing the environment and the transport coefficients are the intrinsic causes reflecting the responsibility of the system driven by the thermal force.

 In what follows, we would like to introduce  a linear in time source for the background metric. So that even in the absence of hydrodynamics the transport coefficients investigated here retain their essential interpretation: they characterize the rate of entropy production when the equilibrium state is subjected to a slowly varying source. Therefore, it is reasonable to write the linear perturbation with both time- and radial-coordinates dependence: $\delta g_{\mu\nu}=t c_0+ h_{\mu\nu}(r)$ with $c_0$ a source. For instance, we are able to write gauge perturbation
 $A_{(i)x}=a_i e^{-i\omega t}=a_i+E_i t+\mathcal{O}(t^2)$.

In order to compute the thermoelectric and thermal conductivities, we  need to
consider perturbations with sources for both the electric and the heat currents.
\bea\label{fl}
g_{tx}=t \delta h(r)+h_{tx},~~~
A_{(1)x}=E_1 t+t a_1(r)+\delta A_1,~~~
A_{(2)x}=E_2 t+t a_2(r)+\delta A_2,
\eea
The conserved currents can be written as
\bea\label{electric}
\mathcal{J}_1&=&-\sqrt{\frac{g_{tt}}{g_{rr}}}Z_1(\phi)\bigg(ta'_1+\delta A'_1\bigg)-q_1 g^{xx}\bigg(t \delta h(r)+h_{tx}\bigg),\\
\mathcal{J}_1&=&-\sqrt{\frac{g_{tt}}{g_{rr}}}Z_2(\phi)\bigg(ta'_2+\delta A'_2\bigg)-q_2 g^{xx}\bigg(t \delta h(r)+h_{tx}\bigg).
\eea
The conserved heat current becomes
\be \label{heat}
\tilde{\mathcal{Q}}=\sqrt{\frac{g_{tt}}{g_{rr}}}\bigg[-g^{tt}\bigg(h_{tx}+t\delta h(r)\bigg)\partial_r g_{tt}+\bigg(t\delta h'(r)+h'_{tx}\bigg)\bigg]-A_{(1)t} J_1-A_{(2)t} J_2.
\ee
In order to evaluate the thermoelectric conductivities, we assume $\delta h(r)=- \zeta g_{tt}$ and $a_i(r)=-E_i+ \zeta A_{(i)t}$, so that the time-dependent terms of the conserved currents
are canceled and the form of the currents remain unchanged.  According to the AdS/CFT dictionary, the coefficient $ \zeta$ corresponds to the thermal
gradient $-\nabla_x T/T$. We can then express the conserved currents (\ref{electric}) and (\ref{heat}) as
\bea
\mathcal{J}_1&=&-\sqrt{\frac{g_{tt}}{g_{rr}}}Z_1(\phi)\delta A'_1-q_1g^{xx}h_{tx},\\
\mathcal{J}_2&=&-\sqrt{\frac{g_{tt}}{g_{rr}}}Z_2(\phi)\delta A'_2-q_2g^{xx}h_{tx},\\
\tilde{\mathcal{Q}}&=&\sqrt{\frac{g_{tt}}{g_{rr}}}\bigg(-g^{tt}h_{tx}\partial_r g_{tt}+h'_{tx}\bigg)-A_{(1)t} J_1-A_{(2)t} J_2.
\eea
In the previous section, we choose the gauge $h_{rx}=0$. Here we would like to turn on $h_{rx}$.
The linearized $rx$-component of the Einstein equations  now is given by
\be
h_{rx}=\frac{g_{xx} \delta \chi'_1}{\beta}+\frac{Z_2(\phi)g_{xx}A'_{(2)t} E_2+Z_1(\phi)g_{xx}A'_{(1)t} E_1}{Y(\phi)g_{tt}\beta^2}+\frac{g_{xx}\delta h'(r)-g'_{xx} \delta h(r)}{g_{tt}\beta^2 Y(\phi)}.
\ee
We assume that $\delta \chi'_1$ is analytic at the event horizon and falls off fast at the infinity so that it has no contribution to the boundary value of $h_{rx}$.
After switching to the Eddington-Finklestein coordinates $(v,r)$ with $v=t+\int \sqrt{g_{rr}/g_{tt}}dr$ and  imposing the regularity condition at the event horizon, from (\ref{fl}) we obtain
\bea
&& \delta A_1=E_1 \int  \sqrt{{g_{rr}}/{g_{tt}}}dr,\\
&&\delta A_2=E_2 \int  \sqrt{{g_{rr}}/{g_{tt}}}dr.
\eea
In the Eddington-Finklestein coordinates, we need explore relationship between $h_{tx}$ and $h_{rx}$. The linear perturbative part of the metric can be expressed as
\be
2 h_{tx}dv dx+2 h_{tx}\sqrt{\frac{g_{rr}}{g_{tt}}} dr dx+2 h_{rx}dr dx.
\ee
In order to cancel out the divergence at the event horizon, we need to impose  the condition
\be\label{htx2}
h_{tx}(r=\rh)= -\sqrt{\frac{g_{tt}}{g_{rr}}} h_{rx}\bigg|_{r=\rh}=\bigg(-\frac{E_1 q_1+E_2 q_2}{Y(\phi_H)\beta^2}-\frac{4\pi T \zeta g_{xx}}{Y(\phi_H)\beta^2}\bigg)\bigg|_{r=\rh}.
\ee
 Therefore, the conserved currents can be expressed by their values at the event horizon
\bea
\mathcal{J}_1&=&\bigg(E_1 Z_1(\phi)+\frac{E_1 q^2_1+E_2 q_1 q_2}{\beta^2Y(\phi)g_{xx}}+\frac{4\pi T q_1 \zeta}{\beta^2 Y(\phi)}\bigg)\bigg|_{r=\rh},\\
\mathcal{J}_2&=&\bigg(E_2 Z_2(\phi)+\frac{E_2 q^2_2+E_1 q_1 q_2}{\beta^2Y(\phi)g_{xx}}+\frac{4\pi T q_2 \zeta}{\beta^2 Y(\phi)}\bigg)\bigg|_{r=\rh},\\
\tilde{\mathcal{Q}}&=& \bigg[ \frac{4\pi T q_1 E_1+4\pi T q_2 E_2}{\beta^2 Y(\phi)}+ \frac{16 \pi^2 T^2 \zeta g_{xx} }{\beta^2 Y(\phi)}\bigg]\bigg|_{r=\rh}.
\eea
The electrical conductivity matrix can be written down as
\bea
\sigma_{11}&=&\frac{\partial \mathcal{J}_1}{\partial E_1}=\rh^{\theta-4}+\frac{q^2_1}{\beta^2}\rh^{2z-4},~~~\sigma_{12}=\frac{\partial \mathcal{J}_1}{\partial E_2}=\frac{q_1 q_2}{\beta^2}\rh^{2z-4},\\
\sigma_{21}&=&\frac{\partial \mathcal{J}_2}{\partial E_1}=\sigma_{12},~~~~~~\sigma_{22}=\frac{\partial \mathcal{J}_2}{\partial E_2}=\rh^{2z-2-\theta}+\frac{q^2_2}{\beta^2}\rh^{2z-4}.
\eea
Therefore, we reproduce the result presented in (\ref{contensor1}).
The thermoelectric conductivity $\alpha$ and thermal conductivity $\bar{\kappa}$ are then evaluated as
\bea
&&\alpha_{1}=\frac{1}{T}\frac{\partial \mathcal{J}_1}{\partial \zeta}=\frac{4\pi q_1}{\beta^2 Y(\phi)}\bigg|_{r=\rh}=\frac{4\pi q_1}{\beta^2 }\rh^{2z-2-\theta},\\
&&\alpha_{2}=\frac{1}{T}\frac{\partial \mathcal{J}_2}{\partial \zeta}=\frac{4\pi q_2}{\beta^2 Y(\phi)}\bigg|_{r=\rh}=\frac{4\pi q_2}{\beta^2 }\rh^{2z-2-\theta},\\
&&\bar{\kappa}=\frac{1}{T}\frac{\partial \tilde{\mathcal{Q}}}{\partial \zeta}=\frac{16 \pi^2 T g_{xx}}{\beta^2 Y(\phi)}\bigg|_{r=\rh}=\frac{16\pi^2 T}{\beta^2}\rh^{2z-2\theta}.
\eea
The thermal conductivity is not influenced by the  gauge fields and only one component appears at this moment.

One can continue of the analysis given before (\ref{mainresult}) and imposes the condition $J_1=0$.  The DC thermoelectric and thermal conductivities become
\bea
&&\bar{\alpha}_{DC}=\frac{4\pi q_2 }{(\beta^2 \rh^{\theta-2z}+q^2_1)\rh^2}.\\
&&\bar{\kappa}_{DC}=\frac{16\pi^2 T}{\beta^2 \rh^{2\theta-2z}+\rh^{\theta}q^2_1}.
\eea
 As $z=1$, $\theta=1$ and $q_1=0$, the Seebeck coefficient behaves as $\bar{\alpha} \sim 1/T$. On the other hand,
  setting $\beta$ to be zero, one obtains $\bar{\alpha} \sim 1/T^2$.
  At zero temperature, $\bar{\alpha} =8\pi q_2/\beta^4$. In brief, the thermoelectric conductivity is influenced by  temperature and impurities.
It would be interesting to check the Wiedemann-Franz law by introducing the thermal conductivity at
zero electric current, which is the usual thermal conductivity that is more readily measurable, $\kappa=\bar{\kappa}_{DC}-\alpha_{DC} \bar{\alpha}_{DC}T/\sigma_{DC}$, and thus
\be
\kappa_{DC}=\frac{16\pi^2 T \rh^{2z+2-2\theta}}{\beta^2 \rh^2+q^2_2 \rh^{\theta}+q^2_1 \rh^{2+2z-\theta}}.
\ee
In the case $\theta=1$, $z=1$ and thus $q_1=0$, both the ratios $\kappa/T=16\pi^2 /\beta^2$ and $\bar{\kappa}/T=16\pi^2 /\beta^2$ are a constant. This reflects that the thermal conductivity is dominated by impurity scattering.

In conventional metals, the WF law is characterized by the constant Lorenz ratio $L_0$. The WF law asserts that the ratio of the electronic contribution of the thermal conductivity to the electrical conductivity of a conventional metal, is proportional to the temperature. This implies that the ability of the quasiparticles to transport heat is determined by their ability to transport charge so the Lorenz ratio is a constant. In our set-up, the Lorenz ratios are given by
\bea\label{lorenz}
&&\bar{L}\equiv \frac{\bar{\kappa}_{DC}}{\sigma_{DC} T}=\frac{16 \pi^2 \rh^{4-\theta}}{\beta^2 \rh^2+q^2_2 \rh^{\theta}+q^2_1 \rh^{2+2z-\theta}},\\
&& L \equiv \frac{{\kappa}_{DC}}{\sigma_{DC} T}=\frac{16  \pi^2 \rh^{6}(\beta^2 \rh^\theta+q^2_1 \rh^{2z})}{(\beta^2 \rh^{2+\theta}+q^2_2 \rh^{2\theta}+q^2_1 \rh^{2+2z})^2}
\label{lorenz2}
\eea
 At zero temperature with $\theta=1$ and $z=1$, (\ref{lorenz}) and (\ref{lorenz2}) yield $\bar{L}=4\pi^2+4\pi^2 \beta^2/\sqrt{4q^2_2+\beta^4}> L_0$ and $L=4\pi^2 \beta^4/(4q^2_2+\beta^4)+4\pi^2 \beta^2/\sqrt{4q^2_2+\beta^4}$.  Usually, we regard $L$ as the quantity comparable with the experiments. (\ref{lorenz2}) implies that as $\beta \rightarrow 0$ deviations from the Fermi-liquid behavior can be obtained, while $\beta \rightarrow \infty$, so $L=8 \pi^2$ the system shows Fermi-liquid-like behavior. This is quiet different from the behavior of the electric conductivity given in (\ref{dcsig}).\\

$\bullet~~~$ $d+2$-dimensional DC transport coefficients \\
In what follows, we extend our results to the $d+2$-dimensional case
\bea
&&\sigma_{11}=\bigg(g^{\frac{d-2}{2}}_{xx}Z_1(\phi)+\frac{q^2_1}{\beta^2 Y(\phi)g^{d/2}_{xx}}\bigg)\bigg|_{r=\rh}=\rh^{2\theta-2\theta/d-2d}+\frac{q^2_1}{\beta^2}\rh^{2z+\theta-2-d-2\theta/d},\\
&&\sigma_{12}=\sigma_{21}=\frac{q_2 q_1}{\beta^2}\rh^{2z+\theta-2-d-2\theta/d},\\
&&\sigma_{22}=\bigg(g^{\frac{d-2}{2}}_{xx}Z_2(\phi)+\frac{q^2_2}{\beta^2 Y(\phi)g^{d/2}_{xx}}\bigg)\bigg|_{r=\rh}=\rh^{d+2z-\theta-4}+\frac{q^2_2}{\beta^2}\rh^{2z+\theta-2-d-2\theta/d},\\
&&\bar{\alpha}_1=\frac{4\pi q_1}{\beta^2 Y(\phi)}\bigg|_{r=\rh}=\frac{4\pi q_1}{\beta^2 }\rh^{2z-2-2\theta/d},\\
&&\bar{\alpha}_2=\frac{4\pi q_2}{\beta^2 Y(\phi)}\bigg|_{r=\rh}=\frac{4\pi q_2}{\beta^2 }\rh^{2z-2-2\theta/d},\\
&&\bar{\kappa}=\frac{16 \pi^2 T g^{d/2}_{xx}}{\beta^2 Y(\phi)}\bigg|_{r=\rh}=\frac{16\pi^2 T}{\beta^2}\rh^{d+2z-2-\theta-2\theta/d}.
\eea
We would also like to generalize the transport coefficients under the
 the condition $J_1=0$, so that the transport coefficients reduce to diagonal components
\bea
\sigma_{DC}&=&\rh^{d+2z-\theta-4}+\frac{q_2 \rh^{2\theta+2z-d-2\theta/d}}{\beta^2 \rh^{2+\theta}+q^2_1 \rh^{2z+d}},\nonumber\\
\alpha_{DC}&=&\bar{\alpha}_{DC}=\frac{4\pi \rh^{2z+\theta-2\theta/d}}{\beta^2 \rh^{2+\theta}+q^2_1 \rh^{2z+d}},\nonumber\\
\bar{\kappa}_{DC}&=&\frac{16\pi^2 T \rh^{2z+d-2\theta/d}}{\beta^2 \rh^{2+\theta}+q^2_1 \rh^{2z+d}}.\nonumber
\eea
Similarly, we have the thermal conductivity at zero electric current
\be
\kappa_{DC}=\frac{16 \pi^2 T \rh^{3d+2z}}{q^2_2 \rh^{4+3\theta}+(\beta^2 \rh^{2+\theta}+q^2_1 \rh^{d+2z})\rh^{2d+2\theta/d}}.
\ee
The corresponding Lorenz ratio in $(d+2)$-dimensional spacetime are obtained as
\bea
&&\bar{L}\equiv \frac{\bar{\kappa}_{DC}}{\sigma_{DC} T}=\frac{16 \pi^2 \rh^{2d+4+\theta}}{q^2_2 \rh^{4+3\theta}+(\beta^2 \rh^{2+\theta}+q^2_1 \rh^{d+2z})\rh^{2d+2\theta/d}},\\
&& L \equiv \frac{{\kappa}_{DC}}{\sigma_{DC} T}=\frac{16  \pi^2 \rh^{4+4d+\theta+2\theta/d}(\beta^2 \rh^{2+\theta}+q^2_1 \rh^{d+2z})}{(q^2_2 \rh^{4+3\theta}+(\beta^2 \rh^{2+\theta}+q^2_1 \rh^{d+2z})\rh^{2d+2\theta/d})^2}.
\eea
At zero temperature with vanishing charge density $q_i=0$, which is associated with the quantum critical regime.  As $z=1$, the above Lorenz ratios are a constant at zero temperature $\bar{L}=L= {16 \pi ^2 \left(d^2-d (\theta +1)+2 \theta \right)}/{d (d-\theta +1)}$. Note that in the absence of charge density, the electric conductivity is dominated by the particle-hole creation of the boundary field theory. While for non-vanishing charge density, the Lorenz ratios decrease as the chemical potential increases. By contrast, $\beta=0$ and $z=1$ at zero temperature leads to $\bar{L}=8\pi^2/(d-\theta)(d+z-\theta)$ and $L=0$.
In general, the Lorenz ratios become temperature independent when $\theta=d$ regardless of the value of $z$, which in turn corresponds to a vanishing  specific heat.

\section{Discussions and conclusions}

In the previous sections, we did not study  the Hall angle, which we would like to defer to a future publication.  After turning on a magnetic field on the background , one can easily find that
the blacken factor is given by
\be
f(r)=1-{\frac{m}{r^{d+z-\theta}}}+{\frac{Q^2_2}{r^{2(d+z-\theta-1)}}}-{\frac{ \beta^2}{r^{2z-2\theta/d}}}+\frac{B^2 r^{2z-6}}{4(1-\theta/d)(4+2\theta/d-3z)}.
\ee
The Hall angle \footnote{ Consistency of the resulting perturbation equations requires both gauge fields to fluctuate. This in turn leads to some subtleties in the analysis.  In general, two gauge fields with magnetic fields lead to a $4\times 4$ DC electrical conductivity matrix. Here we mainly consider Hall angle generated by the second gauge field in the action (1).} can be evaluated by following \cite{Blake:2014yla}
\be
\theta_H \sim \frac{ B q_2 }{\beta^2 }\rh^{2z+\theta-2-d-2\theta/d}.
\ee
For the case $d=2$, $z=1$ and $\theta=1$ and in the higher temperature limit $\rh \sim T$, we have $\theta_{H}\sim T^{-2}$£¬ which is what observed in cuprates. The transport coefficients in the presence of a magnetic field are studied in \cite{gtww} and comparisons with the experimental phenomenologies are discussed.

In summary, we obtained a new black hole solution in  Lifshitz spacetime with a hyperscaling violating factor.
At zero temperature, the black hole approaches $AdS_2\times R^d$ geometry near the horizon with non-vanishing entropy density. One can reproduce the black hole solution given in \cite{Tarrio} and \cite{alisha} as $\theta=0$ and $\beta=0$, respectively. The black hole solution is different from the one obtained in \cite{zyfan}, where the authors  constructed  a class of Lifshitz spacetimes in five dimensions that carry electric fluxes of a Maxwell field.

We then studied holographic DC thermoelectric conductivities in this model with momentum dissipation. The novel matrix method was introduced to compute the transport coefficients.
Since  two gauge fields are presented, these result in a $2\times 2$ DC electric conductivity matrix. The results cannot recover electric conductivity in Reissner-Nordstr$\rm\ddot{o}$m-AdS background  by simply taking $z \rightarrow 1$, $\beta \rightarrow 0$ and $\theta\rightarrow 0$ limits, although the metric can recover that of Reissner-Nordstr$\rm\ddot{o}$m-AdS  type in these limits. This reflects that once we turn on the gauge perturbation in Lifshitz spacetime, it is not possible to have a continuous limit to the perturbation that is normally considered in Reissner-Nordstr$\rm\ddot{o}$m-AdS background.   When we physically setting the electric current  $J_1$ of the auxiliary gauge field to be vanishing, the components of the conductivity matrix with respect to the auxiliary gauge fields disappear, but mixture between $q_1$ and the transport coefficients can be observed. It is only when we take $z=1$, $q_1$ vanishes. We expect that when we turn on the magnetic field in Lifshitz spacetime, the resulting electric conductivity  should be a $4\times 4$ matrix. More complicated situations would then be observed.  It deserves further investigation on such complication and mixture.

The most intriguing result is that linear and quadratic in temperature resistivity can be realized simultaneously under the condition $z=1$, $d=2$ and $\theta=1$. The exponents taken here agree with the scaling approach provided in \cite{Dimi2015}, but different from \cite{hartnoll15,Philip2016}. We notice that the exponents taken here violates the null energy condition in the bulk. But a careful examination of the local thermodynamic stability and the causal structure of the boundary field theory reveals that it is true that the system  is
locally thermodynamically stable at all temperatures and charges without superluminal signal propagation on the boundary.

This work can be considered as a concrete example realizing what were proposed by Blake and Donos in their paper \cite{Blake:2014yla}. For  the resistivity, at the low temperature, it behaves as the Fermi-liquids, while in the high temperature, it reduces to linear in temperature resistivity same as strange metals.
  We also studied the thermoelectric conductivities and the Lorenz ratios in this paper. Although in the holography, there are no quasiparticles  and thus the system has no relationship
   with real Fermi liquids, the scaling geometries presented here are able to mimic Fermi liquid behavior for certain regime of $q^2_2/\beta^4$ as  shown in (\ref{lorenz2}).

\section*{Acknowledgement} We would like to thank Elias Kiritsis, Hong L$\ddot{u}$, Sang-Jin Sin and John McGreevy for helpful discussions.
Y. Tian would like to thank Prof. Glenn Barnich for his hospitality at Universit¨¦ Libre de Bruxelles and International Solvay Institutes.
XHG was partially supported by NSFC,
China (No.11375110).  Y.T. is partially supported by NSFC with Grant No.11475179 and  the Grant (No. 14DZ2260700) from  Shanghai Key Laboratory of High Temperature Superconductors. SYW was partially supported by the Ministry of Science and Technology (grant no. MOST 104-2811-M-009-068) and National Center for Theoretical Sciences in Taiwan. SFW was supported partially by NSFC
China ( No. 11275120 and No. 11675097).


\begin{thebibliography}{99}

\bibitem{brooks} J. R. Schrieffer, James S. Brooks (Ed.), Handbook of high temperature supercodncutvitity,
Springer Press, 2007

\bibitem{Hartnoll}
S.A.Hartnoll, ``\emph{Lectures on holographic methods for condensed matter physics}, "
Class. Quant. Grav. {\bf 26} (2009) 224002 .
[arXiv:0903.3246[hep-th]]

\bibitem{McGreevy}
J. McGreevy, ``\emph{Holographic duality with a view toward many-body physics},"
Adv.High Energy Phys. 723105 (2010).
[arXiv:0909.0518[hep-th]]

\bibitem{Herzog:2009}
C.~P. Herzog, ``\emph{Lectures on holographic superfluidity and
  superconductivity}," J. Phys. {\bf A 42}  (2009) 343001.   [arXiv:0904.1975 [hep-th]]

\bibitem{gws} X. H. Ge, S. J. Sin and S. F. Wu, \emph{Lower Bound of Electrical Conductivity from Holography},  [arXiv:1512.01917 [hep-th]]

\bibitem{Blake:2014yla}
M.~Blake and A.~Donos, \emph{Quantum Critical Transport and the Hall Angle}, Phys. Rev. Lett. {\bf 114}  (2015)  021601
 [arXiv:1406.1659 [hep-th]]

 \bibitem{blaise2} R. A. Davison and B. Gouteraux, \emph{Dissecting holographic conductivities},
   JHEP 1509 (2015)   090 [arXiv:1505.05092[hep-th]].

   \bibitem{blake15} M. Blake, \emph{Momentum relaxation from the fluid/gravity correspondence}, JHEP 09 (2015) 010 [arXiv:1505.06992[hep-th]].

 \bibitem{jianpin} Z. Zhou, J. P. Wu and Y. Ling, \emph{Note on DC and Hall conductivity in holographic massive Einstein-Maxwell-Dilaton gravity},
 JHEP 08 (2015)  067  [arXiv:1504.00535[hep-th]].

\bibitem{pal11}  S. S. Pal, \emph{ Model building in AdS/CMT: DC conductivity and Hall angle}, Phys.Rev. D 84 (2011)
126009, [arXiv:1011.3117].

 \bibitem{pal12}S. S. Pal, \emph{Approximate strange metallic behavior in AdS}, [arXiv:1202.3555].
\bibitem{gout11} B. Gouteraux, B. S. Kim, R. Meyer, \emph{Charged Dilatonic Black Holes and their Transport
Properties}, Fortsch. Phys. 59 (2011) 723-729, [arXiv:1102.4440].
 \bibitem{gout10}B. S. Kim, E. Kiritsis, C. Panagopoulos, \emph{ Holographic quantum criticality and strange metal
transport}, New J. Phys. 14:043045,2012, [arXiv:1012.3464].
 \bibitem{oz13}C. Hoyos, B. S. Kim, Y. Oz, \emph{ Lifshitz Hydrodynamics}, JHEP 1311 (2013) 145, [arXiv:1304.7481].
15
\bibitem{gout14} B. Gouteraux, \emph{Universal scaling properties of extremal cohesive holographic phases}, JHEP 01
(2014) 080, [arXiv:1308.2084].
\bibitem{gout1401} B. Gouteraux, \emph{Charge transport in holography with momentum dissipation}, JHEP 04 (2014)
181, [arXiv:1401.5436].
\bibitem{pang} B. H. Lee, D. W. Pang, C. Park, \emph{A Holographic Model of Strange Metals}, Int. J. Mod. Phys.
A26 (2011) 2279-2305, [arXiv:1107.5822].
\bibitem{pang2}  B. H. Lee, D. W. Pang and C. Park, \emph{Strange Metallic Behavior in Anisotropic Background}, JHEP 1007 (2010)  057 [arXiv:1006.1719 [hep-th]].
\bibitem{lucas} A. Lucas, S. Sachdev,\emph{ Memory matrix theory of magnetotransport in strange metals},
[arXiv:1502.04704].
\bibitem{karch} A. Karch, \emph{Conductivities for Hyperscaling Violating Geometries}, JHEP 1406 (2014) 140,
[arXiv:1405.2926].
 \bibitem{hartnoll}S. A. Hartnoll, A. Karch, \emph{Scaling theory of the cuprate strange metals}, Phys. Rev. B 91
(2015)  155126 [arXiv:1501.03165].
\bibitem{Amoretti16} A. Amoretti, M. Baggioli, N. Magnoli and D. Musso, \emph{Chasing the cuprates with dilatonic dyons}, [arX-
iv:1603.03029 [hep-th]]
\bibitem{kim}
K. Y. Kim, K. K. Kim, Y. Seo and S.J. Sin, \emph{Coherent/incoherent metal transition in a holographic model},
JHEP 1412 (2014)  170 [arXiv:1409.8346[hep-th]]
\bibitem{cheng} L. Cheng, X. H. Ge and Z. Y. Sun, \emph{Thermoelectric DC conductivities with momentum dissipation from higher derivative gravity}, JHEP 1504 (2015) 135 [arXiv:1411.5452[hep-th]].
\bibitem{kim2} K. Y. Kim, K. K. Kim and M. Park, \emph{Ward Identity and Homes Law in a Holographic Superconductor with Momentum Relaxation },  [arXiv:1604.06205 [hep-th]]
\bibitem{tian} Y. Du, C. Niu, Y. Tian and H. B. Zhang, \emph{Holographic thermal relaxation in superfluid turbulence}, JHEP 1512 (2015) 018 [ arXiv:1412.8417 [hep-th]]
\bibitem{andrade} T. Andrade, \emph{ A simple model of momentum relaxation in Lifshtiz holography}, [ arXiv:1602.00556 [hep-th]]
\bibitem{donos} A. Donos and J. P. Gauntlett, \emph{Thermoelectric DC conductivities from black hole horizons}, JHEP 11 (2014) 081 [arXiv:1406.4742]

\bibitem{Horowitz:201204}
G.~T.~Horowitz, J.~E.~Santos and D.~Tong, \emph{Optical Conductivity with Holographic Lattices}, JHEP  1207 (2012) 168 [arXiv:1204.0519 [hep-th]].


\bibitem{Horowitz:2013}
G. T. Horowitz and J. E. Santos, \emph{General Relativity and the Cuprates}, arXiv:1302.6586 [hep-th].


\bibitem{Hartnoll:2012}
A. Donos, S. A. Hartnoll, \emph{Metal-insulator transition in holography}, [arXiv:1212.2998 [hep-th]].

\bibitem{johanna} J. Erdmenger, X. H. Ge and D. W. Pang, \emph{Striped phases in the holographic insulator/superconductor transition,}  JHEP 1311 (2013) 027 [arXiv:1307.4609]

\bibitem{YiLing:2013}
Y. Ling, P. Liu, C. Niu and J. P. Wu, The pseudo-gap phase and the duality in holographic fermionic system with dipole coupling on Q-lattice, Chin.Phys. C40 (2016) no.4, 043102    [arXiv:1602.06062 [hep-th]].

\bibitem{lingprl}
 Y.~Ling, C.~Niu, J.~Wu, Z.~Xian and H.~Zhang,
  ``Metal-insulator Transition by Holographic Charge Density Waves,'' Phys. Rev. Lett. 113 (2014) 091602
  [arXiv:1404.0777 [hep-th]].


\bibitem{Donos:2013eha}
A.~Donos and J.~P. Gauntlett, \emph{Holographic Q-lattices},
 JHEP 1404(2014) 040,
[arXiv:1311.3292 [hep-th]]

\bibitem{YiLing:2014}
Y. Ling, P. Liu, C. Niu, J. P. Wu and Z.-Y. Xian, \emph{Holographic Superconductor on Q-lattice}, JHEP 02 (2015) 059 [arXiv:1410.6761 [hep-th]].

\bibitem{zhou}
Z. Zhou, Y. Ling and J. P. Wu,
\emph{Holographic incoherent transport in Einstein-Maxwell-Dilaton Gravity}, arXiv:1512.01434 [hep-th]


\bibitem{Vegh}
D. Vegh, \emph{Holography without translational symmetry}, arXiv:1301.0537[hep-th].



\bibitem{Davison1}
R.~A. Davison, \emph{Momentum relaxation in holographic massive gravity},
 Phys.Rev. D88 (2013) 086003,
[arXiv:1306.5792 [hep-th]].


\bibitem{Blake:2013}
M.~Blake and D.~Tong, \emph{Universal Resistivity from Holographic Massive
  Gravity},
  Phys.Rev D88 (2013) 106004,
 [arXiv:1308.4970 [hep-th]].

\bibitem{Blake:20132}
M.~Blake, D.~Tong, and D.~Vegh, \emph{Holographic Lattices Give the Graviton a
  Mass},
  Phys.Rev.Lett. 112(2014) 071602,
 [arXiv:1310.3832 [hep-th]].

\bibitem{WJP:2014}
Hua Bi Zeng, Jian-Pin Wu, \emph{Holographic superconductors from the massive gravity},
  Phys. Rev. D 90 (2014) 046001 ,
 [arXiv:1404.5321 [hep-th]].


\bibitem{Davison2}
R.~A. Davison, K.~Schalm, and J.~Zaanen, \emph{Holographic duality and the
  resistivity of strange metals},
   Phys. Rev. B 89(2014) 245116,
 [arXiv:1311.2451 [hep-th]]



\bibitem{withers57}
T. Andrade and B. Withers, \emph{A simple holographic model of momentum
relaxation,}  JHEP 05 (2014) 101, [arXiv:1311.5157 [hep-th]].


\bibitem{Davison:2014}
A. Donos, B. Gout¨¦raux and  E. Kiritsis, \emph{Holographic metals and insulators with helical symmetry}, JHEP 09 (2014) 038  [arXiv:1406.6351]

\bibitem{Gout¨¦raux:2014}
R. A. Davison, B. Gout¨¦raux, \emph{Momentum dissipation and effective theories of coherent and incoherent transport}, JHEP 1501 (2015) 039, arXiv:1411.1062[hep-th]
\bibitem{dibakar} D. Roychowdhury, \emph{Magnetoconductivity in chiral Lifshitz hydrodynamics}, JHEP 09(2015)145 [ arXiv:1508.02002 [hep-th]]

\bibitem{gls} X. H. Ge, Y. Ling, C. Niu and S. J. Sin,\emph{ Thermoelectric conductivities, shear viscosity, and stability in an anisotropic linear axion model },
Phys. Rev. D 92 (2015) 106005 [arXiv:1412.8346[hep-th]].
\bibitem{gautlett1} A. Donos and J. P. Gauntlett, \emph{Navier-Stokes Equations on Black Hole Horizons and DC Thermoelectric Conductivity},
 Phys. Rev. D 92 (2015)  121901.
 \bibitem{gautlett2} E. Banks,  A. Donos and J. P. Gauntlett, \emph{Thermoelectric DC conductivities and Stokes flows on black hole horizons},
 JHEP 1510 (2015)  103.
 \bibitem{gautlett3}  A. Donos, J. P. Gauntlett, T. Griffin and L. Melgar, \emph{DC Conductivity of Magnetised Holographic Matter}, JHEP 01 (2016) 113
 [arXiv:1511.00713]

 \bibitem{blaise3} R. A. Davison, B. Gouteraux and S. Hartnoll, \emph{Incoherent transport in clean quantum critical metals},
   JHEP 1510(2015) 112.

   \bibitem{Amoretti14} A. Amoretti, A. Braggio, N. Maggiore, N. Magnoli and
D. Musso, \emph{Thermo-electric transport in gauge/gravity
models with momentum dissipation}, JHEP 1409
(2014) 160.
\bibitem{Amoretti15} A. Amoretti, A. Braggio, N. Maggiore, N. Magnoli and
D. Musso, \emph{Analytic DC thermo-electric conductivities
in holography with massive gravitons}, Phys. Rev. D 91
(2015) 025002 .
\bibitem{sun} J. R. Sun, S. Y. Wu and H. Q. Zhang,  \emph{Novel Features of the Transport Coefficients in Lifshitz Black Branes}, Phys. Rev. {\bf D 87} (2013) 086005 .
\bibitem{boer}
J. de Boer, K. Papadodimas and E. Verlinde, \emph{Holographic Neutron stars}, [arXiv:0907.2695[hep-th]].
\bibitem{tavanfar}
S. A. Hartnoll and A. Tavanfar,  \emph{Electron stars for holographic metallic criticality}, Phys. Rev. {\bf D 83} (2011) 046003.
\bibitem{allais} A. Allais and J. McGreevy, \emph{How to construct a graviting quantum electron star}, Phys. Rev. {\bf D 88} (2013) 066006.
 \bibitem{alisha} M. Alishahiha, E. O. Colgain and H. Yavartanoo, \emph{Charged black branes with hyperscaling violating factor}, JHEP 1211 (2012) 137 [arXiv:1209.3946[hep-th]]
\bibitem{Tarrio}
J. Tarrio, S. Vandoren,  \emph{Black holes and black branes in Lifshitz spacetimes}, JHEP 1109 (2011) 017
[arXiv:1105.6335[hep-th]]
\bibitem{kuang}X. M. Kuang and J. P. Wu,
\emph{Transport coefficients from hyperscaling violating black brane: shear viscosity and conductivity},  arXiv:1511.03008 [hep-th].
\bibitem{Hlu} S. Cremonini, H. S. Liu, H. L$\ddot{u}$ and C. N. Pope,  \emph{DC conductivities from non-relativistic scaling geoemtries with momentum dissipation},  arXiv:1608.04394 [hep-th].
  \bibitem{liu} N. Iqbal and H. Liu, \emph{Universality of the hydrodynamic limit in AdS/CFT and the membrane paradigm}, Phys. Rev. {\bf D 79} (2009) 025023 [arXiv:0809.3808[hep-th]].
 \bibitem{sonner} J. Sonner,  \emph{On universality of charge transport in AdS/CFT}, JHEP 1307   (2013) 145 [arXiv:1304.7774[hep-th]].
\bibitem{gtww} X. H. Ge, Y. Tian, S. Y. Wu and S. F. Wu,  \emph{Anomalous transport of the cuprate strange metal from holography}, [arXiv£º1606. 05959 [hep-th]].
\bibitem{zyfan} Z. Y. Fan and H. L$\ddot{u}$, \emph{Electrically-charged Lifshitz spacetimes and hyperscaling violations}, JHEP  1504 (2015) 139 [arXiv:1501.05318[hep-th]].
\bibitem{Dimi2015} D. V. Khveshchenko,\emph{ Viable phenomenologies of the normal state of cuprates},  EPL 111 (2015) 1700.
\bibitem{hartnoll15} S. A. Hartnoll and A. Karch, \emph{Scaling theory of the cuprate strange metal},
Phys. Rev. B {\bf 91} (2015) 155126. 
\bibitem{Philip2016} A. Karch, K. Limtragool, P. W. Phillips, \emph{Unparticles and anomalous dimensions in the cuprates}, JHEP {\bf 1603} (2016)  175.



\end{thebibliography}
 \end{document}